\documentclass[12pt]{article}

\newcommand{\beq}{\begin{equation}}
\newcommand{\eeq}{\end{equation}}
\newcommand{\beqa}{\begin{eqnarray}}
\newcommand{\eeqa}{\end{eqnarray}}
\newcommand{\beqar}{\begin{eqnarray*}}
\newcommand{\eeqar}{\end{eqnarray*}}

\newcommand{\inn}{\!\cdot\!}

\newcommand{\la}{\lambda}

\newcommand{\z}{\zeta}

\newcommand{\eg}{{\it e.g.,}\ }
\newcommand{\ie}{{\it i.e.,}\ }
\newcommand{\labell}[1]{\label{#1}} 
\newcommand{\reef}[1]{(\ref{#1})}
\newcommand\prt{\partial}
\newcommand\veps{\varepsilon}
\newcommand{\pol}{\varepsilon}

\newcommand\cR{{\cal R}}
\newcommand\cL{{\cal L}}

\newcommand\cO{{\cal O}}

\newcommand\hR{\hat{R}}

\newcommand\ta{{\tilde a}}
\newcommand\tb{{\tilde b}}
\newcommand\tc{{\tilde c}}
\newcommand\td{{\tilde d}}

\newcommand\tk{{\tilde k}}
\newcommand\ti{{\tilde i}}
\newcommand\tj{{\tilde j}}

\newcommand\Tr{{\rm Tr}}
\newcommand\tr{{\rm tr}}

\begin{document}

\begin{titlepage}

\begin{center}



\vskip 2 cm
{\LARGE \b T-duality  of curvature terms \\
 in D-brane actions}\\
\vskip 1.25 cm
 Mohammad R. Garousi\footnote{garousi@mail.ipm.ir}  \\
 \vskip 1cm
\vskip 1 cm
{{\it Department of Physics, Ferdowsi University of Mashhad\\}{\it P.O. Box 1436, Mashhad, Iran}\\}
\vskip .1 cm
\vskip .1 cm

\end{center}

\vskip 0.5 cm

\begin{abstract}
\baselineskip=18pt
We examine    the known curvature  terms in the DBI part of the D-brane action under  the
 T-duality transformation. Using the  compatibility of the action with the standard rules of T-duality at the linear order as a guiding principle, we include the appropriate  NS-NS $B$-field terms in the action and show that they reproduce the $O(\alpha'^2)$ terms of the corresponding disk-level scattering amplitude. 

\end{abstract}
\vskip 3 cm
\begin{center}
\end{center}
\end{titlepage}
\section{Introduction}
The dynamics of D-branes is well-approximated by the effective world-volume field theories which consist of the sum of Chern-Simons (CS)  and Dirac-Born-Infeld (DBI) actions. The CS part describes the coupling of D-branes to the RR fields which can be found by requiring that the chiral anomaly on the world volume of intersecting D-branes (I-brane) cancels with the anomalous variation of the CS action. This  action for a single D-brane is given by \cite{Green:1996dd,Cheung:1997az,Minasian:1997mm},
\beqa
S_{CS}&=&T_{p}\int_{M^{p+1}}C\wedge(e^{2\pi\alpha'F+B})\wedge\left(\frac{{\cal A}(4\pi^2\alpha'R_T)}{{\cal A}(4\pi^2\alpha'R_N)}\right)^{1/2}\labell{CS}
\eeqa
where $M^{p+1}$ represents the world volume of the D$_p$-brane, $C$ is meant to represent a sum over all appropriate RR potential and ${\cal A}(R_{T,N})$ is the Dirac roof genus of the tangent and normal bundle curvatures respectively,
\beqa
\sqrt{\frac{{\cal A}(4\pi^2\alpha'R_T)}{{\cal A}(4\pi^2\alpha'R_N)}}&=&1+\frac{(4\pi^2\alpha')^2}{384\pi^2}(\tr R_T^2-\tr R_N^2)+\cdots \labell{roof}
\eeqa
For totally-geodesic embeddings of world-volume in the ambient spacetime,   ${\rm R}_{T,N}$ are the pulled back curvature 2-forms of the tangent and normal bundles respectively (see the appendix in ref.
\cite{Bachas:1999um} for more details).

The DBI action on the other hand describes the dynamics of the brane in the presence of the NSNS background fields. At the lowest order it can be found by requiring the consistency with  T-duality \cite{Bachas:1995kx}
\beqa
S_{DBI}&=&-T_p\int d^{p+1}x\,e^{-\phi}\sqrt{-\det\left(G_{ab}+B_{ab}+2\pi\alpha'F_{ab}\right)}
\eeqa
where $G_{ab}$ and $B_{ab}$ are  the pulled back of the bulk fields $G_{\mu\nu}$ and $B_{\mu\nu}$ onto the world-volume of D-brane\footnote{Our index conversion is that the Greek letters  $(\mu,\nu,\cdots)$ are  the indices of the space-time coordinates, the Latin letters $(a,d,c,\cdots)$ are the world-volume indices and the letters $(i,j,k,\cdots)$ are the normal bundle indices.}. The curvature corrections to this action has been found in \cite{Bachas:1999um} by requiring consistency of the effective action with the $O(\alpha'^2)$ terms of the corresponding disk-level scattering amplitude \cite{Garousi:1996ad,Hashimoto:1996kf}. For totally-geodesic embeddings of world-volume in the ambient spacetime, the corrections in string frame  for zero $B_{\mu\nu},F_{ab}$ and for constant dilaton  are \cite{Bachas:1999um}
\beqa
S&\!\!\!\!=\!\!\!\!&\frac{T_p}{4}\frac{(4\pi^2\alpha')^2}{32\pi^2}\int d^{p+1}x\,e^{-\phi}\sqrt{-G}\left(R_{abcd}R^{abcd}-2\hR_{ab}\hR^{ab}-R_{abij}R^{abij}+2\hR_{ij}\hR^{ij}\right)\labell{DBI}
\eeqa
where $\hR_{ab}=G^{cd}R_{cadb}$ and $\hR_{ij}=G^{cd}R_{cidj}$. Here also a tensor with the world-volume or transverse space indices is the pulled back of the corresponding  bulk tensor onto world-volume or transverse space. For the case of D$_3$-brane with trivial normal bundle the above curvature couplings have been modified in \cite{Bachas:1999um} to include the complete sum of D-instanton corrections by  requirement the $SL(2,Z)$ invariance of the couplings.

It has been pointed out in \cite{Myers:1999ps} that the anomalous CS couplings \reef{CS} must be incomplete as they are not compatible with T-duality. T-duality exchanges the components of the metric and  the B-field whereas  the couplings \reef{CS} includes only the curvature terms. 
The same argument can be applied to  the curvature terms in the DBI action to conclude that this action is incompatible  with the T-duality.  However, we will show that in the absence of the B-field the particular combinations of the curvature terms in \reef{DBI} is invariant under T-duality transformation. In this paper we would like to modify the curvature terms in the DBI part to be compatible with the T-duality in the presence of B-field,  at least to linear order in the transformations. 

The constant dilaton appears in the string frame action \reef{DBI} as the overall factor $e^{-\phi}$. The combination $e^{-\phi}\sqrt{-G}$ is invariant under T-duality transformation (see \eg \cite{Myers:1999ps}). When dilaton is non-constant one may expect that the effective action includes the derivatives of $\phi$. These terms however must appear in the action, along with the graviton and B-field,  in a combination of T-dual invariant terms. We will see that it is hard to find a combination of terms involving the derivatives of dilaton which are invariant under the T-duality transformation. Another possibility  is that the dilaton may  appear in the string frame action only as $e^{-\phi}\cL$ where $\cL$ does not include the dilaton. The action in the Einstein frame then includes derivatives of the dilaton through the relation $G_{\mu\nu}=e^{\phi/2}G^E_{\mu\nu}$ which relates the string frame metric to the Einstein frame metric. We will show explicitly by scattering amplitude calculations that the couplings of one dilaton and one graviton, and the couplings of two dilatons at order $\alpha'^2$ appears in the Einstein frame action only through this replacement in the string frame action \reef{DBI}.

An outline of the paper is as follows: We begin the section 2 by introducing our strategy for finding T-dual completion of a term in the effective action. Using this, we find the T-dual completion of the curvature terms in \reef{DBI} in subsections 2.1-2.4. In sections 3 we show that the B-field couplings that we have found are fully consistent with the S-matrix element of two B-fields at order $\alpha'^2$, and write the result in terms of the field strength of B-field, \ie $H=dB$. In section 4 we show that the non-constant dilaton appears in the string frame action only through the overall factor of $e^{-\phi}$. 

\section{T-duality}

In this section  we will show that  the curvature  terms in \reef{DBI}   are inconsistent with T-duality and we will construct a form of the couplings which is consistent with T-duality, at least to linear order in the transformations. 
To simplify the calculations, we are employing static gauge. That is, first we employ spacetime diffeomorphisms to define the D$_p$-brane world-volume as $x^i=0$ with $i=p+1,\cdots, 9$, and then with the world-volume diffeomorphisms, we match the internal coordinates with the remaining spacetime coordinates on the surface, $\sigma^a=x^a$ with $a=0,1,\cdots, p$. In static gauge
the pull-back operations can be ignored and only the restriction of the Riemann tensor to the appropriate subspace is required. We will consider perturbations around flat space where the metric takes the form  $g_{\mu\nu}=\eta_{\mu\nu}+h_{\mu\nu}$ where $h_{\mu\nu}$ is a small perturbation. We denote the Riemann tensor to linear order in $h$ by $\cR_{\mu\nu\rho\lambda}$. This linear Riemann tensor is,
\beqa
\cR_{\mu\nu\rho\la}&=&\frac{1}{2}(h_{\mu\la,\nu\rho}+h_{\nu\rho,\mu\la}-h_{\mu\rho,\nu\la}-h_{\nu\la,\mu\rho})
\eeqa
where as usual commas denote partial differentiation. 

The full set of T-duality transformations has been found in \cite{TB,Meessen:1998qm}.  Assuming the NSNS fields are small perturbations around the flat space and assuming that we are performing T-duality in a Killing direction specified by the label $y$, the relevant transformations take the form (see \eg \cite{Taylor:1999pr}),
\beqa
\tilde{\phi}=\phi-\frac{1}{2}{h}_{yy},\,
\tilde{h}_{yy}=-h_{yy},\, \tilde{h}_{\mu y}=B_{\mu y},\, \tilde{B}_{\mu y}=h_{\mu y},\,\tilde{h}_{\mu\nu}=h_{\mu\nu},\,\tilde{B}_{\mu\nu}=B_{\mu\nu}\labell{Tdual}
\eeqa
 with $\mu,\nu\ne y$. In above transformation the metric is the string frame metric.
 
 Our strategy for finding T-duality invariant couplings corresponding to the curvature terms in \reef{DBI} is as follows: 
Suppose we are implementing T-duality along a world volume direction of D$_p$-brane denoted $y$. A world-volume index in  the contracted Riemann tensors can be separated into  $y$ index and the world-volume index which does not include $y$. We denote this type of  world-volume indices as "tilde" indices, \ie $(\ta,\tb,\tc,\cdots)$. These  indices  are  complete world-volume indices   of the  D$_{p-1}$-brane in  the T-dual theory. However, the $y$ index in the T-dual theory which is a normal bundle index is not a complete index. 
On the other hand, the normal bundle indices of  the original theory \ie $(i,j,k,\cdots)$  are not complete in the T-dual theory. They are not include $y$. We denote this type of normal bundle indices  as ``tilde" indices, \ie $(\ti,\tj,\tk,\cdots )$.  In a T-dual invariant theory, the $y$ indices  must be combined  with the incomplete normal bundle indices $(\ti,\tj,\tk,\cdots )$ to give the complete normal bundle indices  $(i,j,k,\cdots)$ in the T-dual theory.  So  one must add  new terms to the curvature terms of \reef{DBI}  to have   the complete indices in the T-dual theory. 

One may also  implement T-duality along a transverse direction of D$_p$-brane denoted $y$. In this case one must separate the transverse indices to $y$ and the transverse indices $(\ti,\tj,\tk,\cdots )$ which do not include $y$. The  complete world-volume indices of the D$_p$-brane   in the original theory, \ie $(a,b,c,\cdots)$, are not the complete indices of D$_{p+1}$-brane in the T-dual theory which we denote them  $(\ta,\tb,\ta,\cdots)$. They must  include $y$ to be complete.  On the other hand, this time  the ``tilde" transverse indices  are complete indices   of the D$_{p+1}$-brane in the T-dual theory, whereas the $y$ index which is a world-volume index in the T-dual theory must be combined with the incomplete world-volume indices  $(\ta,\tb,\ta,\cdots)$ to give the complete world-volume indices of the D$_{p+1}$-brane. 

Let us consider each curvature terms in \reef{DBI} separately.

\subsection{$\cR_{abcd}\cR^{abcd}$ term}

We begin by examining   the curvature term  $\cR_{abcd}\cR^{abcd}$  under linear T-duality transformation. We are implementing T-duality along a world volume direction of D$_p$-brane. So we write it as
\beqa
(\cR_{abcd})^2&=&(\cR_{\ta\tb\tc\td})^2+(h_{\ta y,\tb\tc}-h_{\tb y,\ta\tc})^2+(h_{yy,\ta\tb})^2
\labell{first0}
\eeqa
Our notation is such that \eg $(\cR_{abcd})^2=\cR_{abcd}\cR^{abcd}$. Under the T-duality transformation \reef{Tdual}, it transforms to 
\beqa
(\cR_{abcd})^2&\rightarrow&(\cR_{\ta\tb\tc\td})^2+(B_{\ta y,\tb\tc}-B_{\tb y,\ta\tc})^2+(h_{yy,\ta\tb})^2
\labell{first00}
\eeqa
Because there are incomplete transverse index $y$, one concludes that the original curvature term is not consistent with  T-duality even in the absence of the B-field. One must add some terms to the curvature term to have completed indices in the T-dual theory. A  T-dual expression  can be constructed from graviton and B-field, or can be constructed from graviton, B-field and dilaton. In the former case, one may  consider the following terms:
\beqa
(\cR_{abcd})^2+(B_{a i,bc}-B_{b i,ac})^2 +\alpha_1(h_{ij,ab})^2+\beta_1(B_{ci,ab})^2+\gamma_1(h_{cd,ab})^2\nonumber
\eeqa
where $ \alpha_1,\, \beta_1$ and $\gamma_1$ are three  constants that we will find by requiring the above combination to be invariant under the T-duality transformation \reef{Tdual}. Doing the same steps as we have done for $(\cR_{abcd})^2$, one finds that the above expression transforms under the T-duality to the following terms:
\beqa
(h_{yy,\ta\tb})^2(-1+\alpha_1 -\beta_1+\gamma_1)+(h_{yi,\ta\ta})^2(2 -2\alpha_1+\beta_1)+(B_{y\tc,\ta\tb})^2(-\beta_1+2\gamma_1)+\cdots
\labell{one}
\eeqa
where dots refer to the terms which have complete indices.  

To have a T-dual expression, one must  impose the vanishing of the coefficient of the above three terms which have the incomplete index $y$. Since the three equations are not independent this fixes $ \beta_1$ and $\gamma_1$. The T-dual completion of the curvature term is then
\beqa
(\cR_{abcd})^2+(B_{a i,bc}-B_{b i,ac})^2+\alpha_1(h_{ij,ab})^2-(2-2\alpha_1)(B_{ci,ab})^2-(1-\alpha_1)(h_{cd,ab})^2
\labell{first}
\eeqa
 Because of the third and the  last terms in the above expression which are not tensor under the diffeomorphism, the effective action can not have only the curvature $(\cR_{abcd})^2$, as it is.  We will see that the non-tensor gravity  terms in \reef{first}    disappear in the total  curvature terms in \reef{DBI}. Similarly, the B-field couplings above can not be written  in terms of the field strength of B-field, \ie $H=dB$, which again confirms that the effective action can not have only the curvature $(\cR_{abcd})^2$.  We will see that all B-field couplings corresponding to  the total  curvature terms in \reef{DBI} can be written in terms of $H$.

 One may try to find a T-dual completion of $(\cR_{abcd})^2$ by including the derivative of the  dilaton field. We have found  that it is very unlikely to have such  a T-dual completion. We will consider the effect of the non-constant dilaton in section 4.

\subsection{$\hat{\cR}_{ab}\hat{\cR}^{ab}$ term}

We begin this subsection by noting that the linear curvature $\hat{\cR}_{ab}$ has on-shell ambiguity. The linear form of this curvature as defined in \reef{DBI} is 
\beqa
{\cR}_{cacb}&=&\frac{1}{2}\left(h_{cb,ac}+h_{ac,cb}-h_{cc,ab}-h_{ab,cc}\right)
\eeqa
Using the on-shell conditions that graviton is traceless and satisfies $\prt_{\mu}h_{\mu\nu}=0$, one can write the above equation as
\beqa
-\frac{1}{2}\left(h_{ib,ai}+h_{ai,ib}-h_{ii,ab}-h_{ab,ii}\right)
\eeqa
which is $- {\cR}_{iaib}$. One can easily observe that neither of the above expressions is invariant under T-duality when  $a,b\ne y$. The third term in the first expression  can be decomposed as $h_{\tc\tc,\ta\tb}+h_{yy,\ta\tb}$ and transforms to $h_{\tc\tc,\ta\tb}-h_{yy,\ta\tb}$ under T-duality which is not invariant. Similarly, the third term in the second expression  above  transforms to $h_{\ti\ti,\ta\tb}$ under T-duality which is not  invariant either. However, the combination $\frac{1}{2}(h_{cc,ab}-h_{ii,ab})$ can be decomposed as $\frac{1}{2}(h_{\tc\tc,\ta\tb}+h_{yy,\ta\tb}-h_{ii,\ta\tb})$ which  transforms to $\frac{1}{2}(h_{\tc\tc,\ta\tb}-h_{yy,\ta\tb}-h_{\ti\ti,\ta\tb})$ under T-duality. This can be written as $\frac{1}{2}(h_{\tc\tc,\ta\tb}-h_{ii,\ta\tb})$. So the combination  $\frac{1}{2}(h_{cc,ab}-h_{ii,ab})$ is invariant under the T-duality transformation. We fix the above ambiguity by choosing the combination that is invariant under T-duality, \ie
\beqa
\hat{\cR}_{ab}&=&\frac{1}{2}({\cR}_{cacb}-{\cR}_{iaib})
\eeqa
Using this expression for the curvature, we now find the T-dual completion of the curvature squared. The world-volume indices of this term can be decomposed  as
\beqa
(\hat{\cR}_{ab})^2&=&(\hat{\cR}_{\ta\tb})^2+\frac{1}{2}(h_{\tc y,\ta\tc}-h_{\ta y,\tc\tc})^2+\frac{1}{4}(h_{yy,\tc\tc})^2
\nonumber
\eeqa
Under the T-duality  transformation \reef{Tdual}, it transforms to 
\beqa
(\hat{\cR}_{ab})^2&\rightarrow&(\hat{\cR}_{\ta\tb})^2+\frac{1}{2}(B_{\tc y,\ta\tc}-B_{\ta y,\tc\tc})^2+\frac{1}{4}(h_{yy,\tc\tc})^2
\nonumber
\eeqa
There are incomplete transverse index $y$ in the second and last terms, so one concludes that the original curvature term is not consistent with  T-duality.  One may  consider the following terms:
\beqa
(\hat{\cR}_{ab})^2+\frac{1}{2}(B_{ic ,ca}-B_{ia,cc})^2+\beta_2(B_{ia,cc})^2+\gamma_2(h_{ab,cc})^2
\labell{second1}
\eeqa
where $ \beta_2$ and $\gamma_2$ are two constants.  Doing the same steps as we have done for $(\hat{\cR}_{ab})^2$, one finds that the above expression transforms under the T-duality to the following terms:
\beqa
(h_{yy,\tc\tc})^2(\frac{1}{4}  +\gamma_2)+(h_{yi,\tc\tc})^2(\frac{1}{2} +\beta_2)+(B_{y\ta,\tc\tc})^2(-\beta_2+2\gamma_2)+\cdots
\labell{two}
\eeqa
where dots refer to the terms which have complete indices. To have a T-dual expression, one may  again impose the vanishing of the coefficient of the above three terms which have the incomplete index $y$.    
The T-dual completion in this case is  the following couplings:
\beqa
(\hat{\cR}_{ab})^2+\frac{1}{2}(B_{ic ,ca}-B_{ia,cc})^2-\frac{1}{2}(B_{ia,cc})^2-\frac{1}{4}(h_{ab,cc})^2
\labell{second}
\eeqa
 The curvature term  again is not invariant under T-duality even in the absence of B-field. 

As a cross check of the equations \reef{first} and \reef{second}, we find the T-dual completion  of the Gauss-Bonnet combination at quadratic order. One can easily observes that $\hat{\cR}^2$ where $\hat{\cR}$ is obtained from  $\hat{\cR}_{ab}$ by contracting the tangent indices,  is invariant under T-duality. Using  the equation \reef{first} for $\alpha_1=0$ case and equation  \reef{second}, one finds that the T-dual completion of the Gauss-Bonnet combination at quadratic order is
\beqa
&&(\cR_{abcd})^2-4(\hat{\cR}_{ab})^2+\hat{\cR}^2+\nonumber\\
&&+(B_{ai,bc}-B_{bi,ac})^2-2(B_{ci,ab})^2-2(B_{ic,ca}-B_{ia,cc})^2+2(B_{ia,cc})^2 \nonumber
\eeqa
where we have  ignored some total derivative terms. The first line is the Gauss-Bonnet combination and the second line is its corresponding B-fields. Note that the non-tensor terms cancel each other in the T-dual completion of the Gauss-Bonnet combination. The Gauss-Bonnet combination is a topological invariant in four dimensions, but it is a total derivative at quadratic order in all dimensions. One can easily check that the above B-field terms  can be written as a  total derivative term as expected.
\subsection{$\cR_{abij}\cR^{abij}$ term}

The world-volume indices of this term can be written as
\beqa
(\cR_{abij})^2&=&(\cR_{\ta\tb ij})^2+\frac{1}{2}(h_{j y,i\ta}-h_{i y,j\ta})^2
\nonumber
\eeqa
Under the T-duality  transformation \reef{Tdual}, this transforms to 
\beqa
(\cR_{abij})^2&\rightarrow&(\cR_{\ta\tb \ti\tj})^2+\frac{1}{2}(B_{\tj y,\ti\ta}-B_{\ti y,\tj\ta})^2
\nonumber\\
&&=(\cR_{\ta\tb ij})^2-\frac{1}{2}(h_{\ta y,\tb i}-h_{\tb y,\ta i})^2+\frac{1}{2}(B_{j y,\ti\ta}-B_{i y,\tj\ta})^2
\eeqa
where in the second line we have written the incomplete indices $\ti,\tj$ in terms of the complete indices $i,j$. There is still incomplete transverse index $y$ in the second  and third terms.  So  the original curvature term is not consistent with  T-duality. The T-dual completion in this case is  the following couplings:
\beqa
(\cR_{abij})^2+\frac{1}{2}(B_{ki ,aj}-B_{kj,ai})^2+\frac{1}{2}(B_{ac ,bi}-B_{bc,ai})^2-\frac{1}{2}(B_{ki,aj})^2-\frac{1}{2}(B_{ac,bi})^2
\labell{third}
\eeqa
In this case the curvature term alone  is  invariant under T-duality  in the absence of B-field. 

\subsection{$\hat{\cR}_{ij}\hat{\cR}^{ij}$ term}

This curvature also has on-shell ambiguity. The linear form of this curvature as defined in the  action \reef{DBI} is ${\cR}_{cicj}$, however, using the on-shell condition it can be written as $-{\cR}_{kikj}$. In this case also we fix this ambiguity by choosing the T-dual invariant combination, \ie
\beqa
\hat{\cR}_{ij}&=&\frac{1}{2}({\cR}_{cicj}-{\cR}_{kikj})
\eeqa
Since the curvature $\hat{\cR}_{ij}$ is T-duality invariant when $i,j\ne y$, the transformation of the curvature squared term $(\hat{\cR}_{ij})^2$ under the T-duality becomes
\beqa
(\hat{\cR}_{ij})^2&\rightarrow&(\hat{\cR}_{\ti\tj})^2
\nonumber\\
&&=(\hat{\cR}_{ij})^2-\frac{1}{2}(h_{\ta y,\ta j}-h_{yj,\ta\ta})^2+\frac{1}{4}(h_{yy,\ta\ta})^2\nonumber
\eeqa
There is the incomplete transverse index $y$ on the left hand side of the  above equation, so  the original curvature term is not consistent with  T-duality.  To have a T-dual expression, one should consider the following terms:
\beqa
(\hat{\cR}_{ij})^2+\frac{1}{2}(B_{ab,bj}-B_{aj,bb})^2+\alpha_3(h_{ij,cc})^2+\beta_3(B_{ai,cc})^2 
\labell{fourth1}
\eeqa
where $\alpha_3$ and $\beta_3$   are two constants.  Doing the same steps as before, one finds that the above expression transforms under the T-duality to the following terms:
\beqa
(h_{yy,\tc\tc})^2(\frac{1}{4}+\alpha_3    )+(h_{yi,\tc\tc})^2(-2\alpha_3 +\beta_3)+(B_{y\ta,\tc\tc})^2(-\frac{1}{2}-\beta_3 )+\cdots
\labell{three}
\eeqa
where dots refer to the terms which have complete indices. To have a T-dual expression, one must  again impose the vanishing of the coefficient of the above three terms which have the incomplete index $y$.    
The T-dual completion in this case is  the following couplings:
 
\beqa
(\hat{\cR}_{ij})^2+\frac{1}{2}(B_{ab,bj}-B_{aj,bb})^2-\frac{1}{4}(h_{ij,cc})^2-\frac{1}{2}(B_{ai,cc})^2
\labell{fourth}
\eeqa
which indicates that the curvature term is not invariant under the T-duality transformation \reef{Tdual} even in the absence of the B-field.

The non-covariant gravity term in \reef{fourth}   appears also  in the   T-dual invariant expression \reef{first}. To cancel this term one must fix the constant $\alpha_1$ in \reef{first} to be
\beqa
\alpha_1=\frac{1}{2}
\eeqa
Using this constant,  one finds that the gravity term $(h_{ab,cc})^2$  in the  T-dual expressions \reef{first} and \reef{second} is also canceled.   As a result, the combination of the T-dual expressions \reef{first}, \reef{second} and \reef{fourth} is  covariant.   
\section{$\prt H\prt H$ couplings}

Now adding the equations \reef{first}, \reef{second}, \reef{third},  \reef{fourth}, and using the integration by part, \eg $(h_{ab,cd})^2=(h_{ab,cc})^2$, one finds that the non-tensor graviton terms, \eg $(h_{ab,cd})^2$,  are canceled among  the   curvature terms in \reef{DBI}, \ie the particular combination of the curvature squared terms in \reef{DBI} is invariant under the linear T-duality transformations \reef{Tdual} in the absence of the B-field. In the presence of B-field,   consistency with the T-duality transformations requires that one must include in the action \reef{DBI} the B-field couplings  \reef{first}, \reef{second}, \reef{third},  \reef{fourth}. Ignoring some total derivative terms, one finds the following couplings:
\beqa
&&B_{ki,aj}B_{kj,ai}+B_{ac,bi}B_{bc,ai}-\frac{1}{2}(B_{ki,aj})^2-\frac{1}{2}(B_{ac,bi})^2\nonumber\\
&&-(B_{ic,ca})^2-2B_{ab,bi}B_{ai,cc}+(B_{ab,bi})^2+(B_{ai,bb})^2\labell{2B-field}
\eeqa
To check that these couplings are the complete couplings of two B-fields to the D$_p$-brane, they  must be reproduced by the $\cO(\alpha'^2)$  contact terms of the corresponding disk-level scattering amplitude. The scattering amplitude of two NSNS states from D$_p$-brane  is given by \cite{Garousi:1996ad,Garousi:1998bj}
 \beqa
 A(\veps_1,p_1;\veps_2,p_2)&=&-\frac{1}{8}T_p\alpha'^2K(1,2)\frac{\Gamma(-\alpha' t/4)\Gamma(\alpha' q^2)}{\Gamma(1-\alpha't/4+\alpha' q^2)}\nonumber\\
 &=&\frac{1}{2}T_pK(1,2)\left(\frac{1}{q^2t}+\frac{\pi^2\alpha'^2}{24}+O(\alpha'^4)\right)\labell{Amp}
 \eeqa
where $q^2=p_1^ap_1^b\eta_{ab}$ is the momentum flowing along the world-volume of D-brane, and $t=-(p_1+p_2)^2$ is the momentum transfer in the transverse direction. The kinematic factor is\footnote{Note that the expressions for $a_1,a_2$ in \cite{Garousi:1996ad} are valid only for the symmetric polarizations, whereas the expressions in \cite{Garousi:1998bj} are valid for both symmetric and antisymmetric polarizations.}
\beqa
K(1,2)&=&\left(2q^2a_1+\frac{t}{2}a_2\right)\labell{kin}
\eeqa
where
\beqa
a_1&=&{\rm Tr}(\pol_1\inn D)\,p_1\inn \pol_2 \inn p_1 -p_1\inn\pol_2\inn
D\inn\pol_1\inn p_2 - p_1\inn\pol_2\inn\pol_1^T \inn D\inn p_1
-p_1\inn\pol^T_2\inn\pol_1\inn D\inn p_1
\nonumber\\
&&-\frac{1}{2}(p_1\inn\pol_2\inn \pol^T_1\inn p_2+p_2\inn\pol^T_1\inn\pol_2
\inn p_1)+
q^2\,{\rm Tr}(\pol_1\inn\pol_2^T)
+\Big\{1\longleftrightarrow 2\Big\}
\labell{fintwo}\\
\nonumber\\
a_2&=&{\rm Tr}(\pol_1\inn D)\,(p_1\inn\pol_2\inn D\inn p_2 + p_2\inn
D\inn\pol_2\inn p_1 +p_2\inn D\inn\pol_2\inn D\inn p_2)
+p_1\inn D\inn\pol_1\inn D\inn\pol_2\inn D\inn p_2
\nonumber\\
&&-\frac{1}{2}(p_2\inn
D\inn\pol_2\inn\pol_1^T\inn D\inn p_1+p_1\inn D\inn\pol^T_1\inn\pol_2\inn
D\inn p_2)
+q^2\,{\rm Tr}(\pol_1\inn D\inn \pol_2\inn D)
\nonumber\\
&&-q^2\,{\rm Tr}(\pol_1\inn\pol_2^T)
-{\rm Tr}(\pol_1\inn D) {\rm Tr}(\pol_2\inn D)\,(q^2-t/4)
+\Big\{1\longleftrightarrow 2 \Big\}\ \ ,
\labell{finthree}\nonumber
\eeqa
where the diagonal matrix $D$ has a $+1$ entry for the direction tangent to the world-volume of the static D$_p$-brane and $-1$ for a normal direction.

The leading terms of the expansion \reef{Amp} are reproduced by the field theory consisting of  the DBI action and the two-derivative bulk supergravity action \cite{Garousi:1996ad,Garousi:1998bj}. It has been shown in \cite{Bachas:1999um} that the $\cO(\alpha'^2)$ terms of the above amplitude when the two NSNS states are gravitons are reproduced by the curvature squared terms in \reef{DBI}. We have explicitly check that   the $\cO(\alpha'^2)$ terms of the string amplitude \reef{Amp} when the two NSNS states are B-field are exactly reproduced  by the D-brane couplings in \reef{2B-field}.

We now  try to write the couplings \reef{2B-field} in terms of field strength. In \cite{Gross:1986mw} it has been shown that the four B-fields  corrections to supergravity action can be written in terms of  the generalized $R^4$ terms in which the spin connection is shifted by the field strength of B-field, \ie $\omega\rightarrow \omega+H$ which gives the shift $R_{\mu\nu\rho\lambda}\rightarrow R_{\mu\nu\rho\lambda}+\prt_{[\mu} H_{\nu]\rho\lambda}$. One may expect that the two B-field couplings that we have found can also  be written in terms of the generalized curvature terms. However, there is no such an analogy  in the effective action of the D-brane. To see this we note that the couplings in \reef{2B-field} have either three world-volume and one normal bundle indices or three normal bundle and one world-volume indices, whereas the curvature couplings in \reef{DBI} have either four world-volume indices or two world-volume and two normal bundle indices.  

One can easily show that the terms in \reef{2B-field} which have three normal bundle  and one world-volume indices can be written as $-\frac{1}{6}(\prt_aH_{ijk})^2$ where $H_{ijk}=B_{ij,k}+B_{ki,j}+B_{jk,i}$, and the rest  can be written as$-\frac{1}{3}(\prt_iH_{abc})^2+\frac{1}{2}(\prt_aH_{bci})^2$. So the $\cO(\alpha'^2)$ correction to the DBI action when the NSNS fields are small perturbations around the flat space and when $F_{ab}=0$ is given by the following couplings: 
\beqa
S&=&\frac{T_p}{4}\frac{(4\pi^2\alpha')^2}{32\pi^2}\int d^{p+1}x\,e^{-\phi}\sqrt{-G}\bigg[\cR_{abcd}\cR^{abcd}-2\hat{\cR}_{ab}\hat{\cR}^{ab}-\cR_{abij}\cR^{abij}+2\hat{\cR}_{ij}\hat{\cR}^{ij}\nonumber\\
&&-\frac{1}{6}\prt_aH_{ijk}\prt^aH^{ijk}-\frac{1}{3}\prt_iH_{abc}\prt^iH^{abc}+\frac{1}{2}\prt_aH_{bci}\prt^aH^{bci}
\bigg]\labell{LDBI}
\eeqa
where the indices are raised and lowered by the flat metrics $\eta_{ab}$ and $\eta_{ij}$. So far we have assumed that the dilaton is a constant. In the next section we consider the non-constant dilaton case. It turns out that the above action is valid even for non-constant dilaton field. 

\section{Dilaton couplings}

We have seen that by using the linearized T-duality transformation \reef{Tdual} for graviton and B-field, one finds that the curvature terms in \reef{DBI} are invariant under T-duality. On the other hand the dilaton transforms to graviton under T-duality. If one assumes that there are some terms in the action \reef{DBI} at order $\cO(\alpha'^2)$ which involve the derivative of dilaton, then  the dilaton transformation would produce 
 terms like $\prt\prt\phi\prt\prt h_{yy}$ which ruins the T-duality of the curvature terms in \reef{DBI}. So the consistency with  the T-duality transformations \reef{Tdual} may indicate that there is no dilaton couplings at order $\cO(\alpha'^2)$ in the string frame action. To state it differently, consider  the Einstein frame couplings, \ie $G_{\mu\nu}=e^{\phi/2}G^{E}_{\mu\nu}$ which  is 
\beqa
h_{\mu\nu}&=&h^E_{\mu\nu}+\frac{\phi}{2}\eta_{\mu\nu}
\eeqa
at the linear order. To find  the Einstein frame couplings of $n$  dilatons and $m$ gravitons  from  the corresponding string frame couplings of   $n+m$ gravitons, one must replace $n$  gravitons  by $n$ dilatons, \ie $h_{\mu\nu}h_{\alpha\beta}\cdots\rightarrow\phi\eta_{\mu\nu}\phi\eta_{\alpha\beta}\cdots$. The above speculation is that the derivative of the dilaton  appear in the  Einstein frame action only through this replacement. In the momentum space, this replacement corresponds to replacing a graviton polarization $\veps_{\mu\nu}$ with $\eta_{\mu\nu}$.  

One the other hand, it is known that  the S-matrix elements of dilaton is given by  the S-matrix elements of massless NSNS state in which the polarization is the following:
\beqa
\veps_{\mu\nu}&=&\frac{1}{\sqrt{D-2}}(\eta_{\mu\nu}-\ell_{\mu}p_{\nu}-\ell_{\nu}p_{\mu})\,;\qquad \ell\inn p=1\labell{pol}
\eeqa
where the auxiliary vector  $\ell_{\mu}$ insures that the polarization satisfies the on-shell condition $p\inn\veps_{\nu}=0$.  As an example consider  the S-matrix element of one massless NSNS state and one transverse scalar field in superstrig theory  \cite{Garousi:1998fg}
\beqa
A(\lambda,h)&\sim&4(k_1)_a(\veps_2)^{ai}(\z_1)_i+(p_2)_i(\z_1)^i\Tr(\veps_2\inn D)\labell{amp1}
\eeqa
This amplitude for graviton produces  the following couplings of the DBI action:
\beqa
\cL(\lambda,h)&\sim&2h_{ia}\prt^a\lambda^i+\lambda^i\prt_ih^a{}_a\labell{amp11}
\eeqa
where the first terms is coming from the pull-back of metric $h_{\mu\nu}$ and the last term is Taylor expansion of $h^a{}_a$.
Now if one replaces the polarization  \reef{pol} into the  amplitude \reef{amp1}, one will find
\beqa
A(\lambda,\phi)&\sim&\frac{p-3}{2\sqrt{2}}(\z_1)_ip_2^i\labell{amp2}
\eeqa
which produces  the Taylor expansion of the dilaton in the DBI action in the Einstein frame, \ie
\beqa
\cL(\lambda,\phi)&\sim&\frac{p-3}{2\sqrt{2}}\lambda^i\prt_i\phi\labell{amp22}
\eeqa
The amplitude \reef{amp2} can not be found from the  amplitude \reef{amp1} by replacing the graviton polarization $\veps_{\mu\nu}$ in \reef{amp1} by $\eta_{\mu\nu}$. This may indicate that the dilaton coupling \reef{amp22}  is not given by   the  string frame  couplings \reef{amp11} transformed  to the Einstein frame. However, the difference is the extra factor of  $e^{-\phi}$ in the string frame action. 

Using the above observations, one may speculate that, apart from the extra factor of $e^{-\phi}$, the dilaton couplings in the Einstein frame are given by  the string frame graviton couplings which transformed to the Einstein frame. We will show that this is the case for the couplings of one dilaton and one graviton, and for two dilatons couplings at order $\alpha'^2$. 

The S-matrix element of one graviton and one dilaton is given by \reef{Amp} in which  one of the graviton polarizations, \eg $(\veps_1)_{\mu\nu}$ is replaced by \reef{pol}. The kinematic factor \reef{kin} becomes
\beqa
K(\phi,h)&=&(\Tr(D)+2)\left(\frac{}{}p_1\inn D\inn p_1\,p_1\inn\veps_2\inn p_1-\right.\labell{kin1}\\
&&\left.-p_1\inn p_2[2p_1\inn\veps_2\inn D\inn p_2+p_2\inn D\inn\veps_2\inn D\inn p_2+p_1\inn D\inn p_2\Tr(\veps_2\inn D)]\frac{}{}\right)\nonumber
\eeqa
Note that the auxiliary vector $\ell_{\mu}$ disappeared and the amplitude is zero for the D$_3$-brane, as expected. One can easily check that the above kinematic factor can be obtained from the kinematic factor of two gravitons by replacing  the graviton polarization $(\veps_1)_{\mu\nu}$ with $\eta_{\mu\nu}$. This is consistent with the above speculation. In fact the curvature terms in \reef{DBI} for non-constant dilaton gives the following couplings in the Einstein frame:
\beqa
(\Tr(D)+2)[\hat{\cR}^{ab}\phi_{,ab}-\hat{\cR}^{ij}\phi_{,ij}]
\eeqa
which reproduces exactly the amplitude \reef{kin1}.

Finally, the S-matrix element of two dilatons is given by \reef{Amp} in which both graviton polarizations are replaced by \reef{pol}. The kinematic factor \reef{kin} in this case becomes
\beqa
K(\phi,\phi)&=&-(\Tr(D)+2)^2\,p_1\inn p_2\,p_1\inn D\inn p_2\labell{kin2}
\eeqa
which  can easily be found from the  kinematic factor \reef{kin1}  by replacing  the graviton polarization $(\veps_2)_{\mu\nu}$ with $\eta_{\mu\nu}$. This is again consistent with the above speculation that  the couplings of two dilatons in the Einstein frame are given by the couplings of two gravitons in the string frame  transformed to the Einstein frame. That is, the string frame  curvature terms in \reef{DBI} for non-constant dilaton gives the following couplings in the Einstein frame:
\beqa
(\Tr(D)+2)^2[(\phi_{,ab})^2-(\phi_{,ij})^2]
\eeqa
which reproduces  the dilaton amplitude \reef{kin2}, as expected.

{\bf Acknowledgments}: I would like to thank Katrin Becker and Rob Myers for many useful  discussions.  This work is supported by Ferdowsi University of Mashhad under grant p/2127(88/4/27). 
\bibliographystyle{/Users/Nick/utphys} 
\bibliographystyle{utphys} \bibliography{hyperrefs-final}


\providecommand{\href}[2]{#2}\begingroup\raggedright

\endgroup

\end{document}